\documentclass{iau307}
\usepackage{graphicx}
\usepackage{natbib}
\usepackage{url}
\usepackage{dtklogos}
\bibpunct{(}{)}{;}{a}{}{,}

\title[The puzzling new variable stars in NGC~3766] 
{Are the stars of a new class of variability detected in NGC~3766 fast rotating SPB stars?}

\author[Salmon S.J.A.J. et al.]   
{S.J.A.J. Salmon$^1$
 \and J. Montalb\`an$^1$
 \and D.R. Reese$^3$
 \and M-A. Dupret$^1$
 \and P. Eggenberger$^2$
}

\affiliation{$^1$D\'{e}partement d'Astrophysique, G\'{e}ophysique et Oc\'{e}anographie, Universit\'{e} de Li\`{e}ge, All\'{e}e du 6 Ao\^{u}t 17, 4000 Li\`{e}ge, Belgium \\ email: {\tt salmon@astro.ulg.ac.be} \\[\affilskip]
$^2$Observatoire de Gen\`{e}ve, Universit\'{e} de Gen\`{e}ve, Chemin des Maillettes 51, 1290 Sauverny, Switzerland \\[\affilskip]
$^3$ School of Physics and Astronomy, University of Birmingham, Edgbaston, Birmingham, B15 2TT, UK}

\pubyear{2014}
\volume{307} 
\pagerange{}
\jname{New windows on massive stars: asteroseismology, interferometry, and spectropolarimetry}
\editors{G. Meynet, C. Georgy, J.H. Groh \& Ph. Stee, eds.}

\begin{document}

\maketitle
\begin{abstract}

A recent photometric survey in the NGC~3766 cluster led to the detection of stars presenting an unexpected variability. They lie in a region of the Hertzsprung-Russell (HR) diagram where no pulsation are theoretically expected, in between the $\delta$ Scuti and slowly pulsating B (SPB) star instability domains. Their variability periods, between $\sim$0.1--0.7~d, are outside the expected domains of these well-known pulsators. The NCG~3766 cluster is known to host fast rotating stars. Rotation can significantly affect the pulsation properties of stars and alter their apparent luminosity through gravity darkening. Therefore we inspect if the new variable stars could correspond to fast rotating SPB stars. We carry out instability and visibility analysis of SPB pulsation modes within the frame of the traditional approximation. The effects of gravity darkening on typical SPB models are next studied. We find that at the red border of the SPB instability strip, prograde sectoral (PS) modes are preferentially excited, with periods shifted in the 0.2--0.5~d range due to the Coriolis effect. These modes are best seen when the star is seen equator-on. For such inclinations, low-mass SPB models can appear fainter due to gravity darkening and as if they were located between the $\delta$~Scuti and SPB instability strips.  

\end{abstract}

\firstsection 
\section{Introduction}
\label{SecO}
Intermediate-mass stars can exhibit various types of pulsation during the main sequence; mid- to late B-type stars present high-order gravity (g) modes with periods $\gtrsim$ 1~d (SPB pulsators), whereas late A- and early F-type stars show low-order pressure (p) and g modes with periods between 0.3~h and 6~hr ($\delta$ Scuti stars). These pulsations are driven by the $\kappa$ mechanism due to the iron-group and HeII opacity bumps, respectively. In early A-type stars, none of these two opacity bumps fit the conditions to efficiently activate the $\kappa$ mechanism \citep[e.g.][]{pami99}, so that they are not expected to pulsate.

However, \citet[][hereafter Mo13]{mowlavi} recently detected a significant number (36) of unknown variable stars in NGC~3766. These stars span over 3 magnitudes and are all located in the HR diagram between those identified as SPB or $\delta$~Scuti candidates. The amplitudes of these new variable stars are typically lower by a factor 2 to 3. Moreover, their periods are between $\sim$0.1~d and 0.7~d, distinguishing them from SPB or $\delta$~Scuti modes. The existence of variable stars with such properties, often referred as Maia stars, is a recurrent debate \citep[e.g.][]{decatmaia}. Yet, low-amplitude variable stars observed in the field of the CoRoT mission \citep{degrootecorot} and in NGC~884 \citep{saesenpersei} appeared as possible new such candidates, with properties similar to those found in NGC~3766 by Mo13. Quite recently, \citet{polonais} reported the detection of a very similar population of variable stars in the NGC~457 cluster, rising once again the question about the possible origin of this variability. 


Mo13 already suggested the role of rotation, since B stars of the NGC~3766 cluster are known to be fast rotators with typical rotational velocities half or more their break-up velocities \citep{mcswain}. Here we advance that the new variable stars are in fact fast rotating SPBs. Using the traditional approximation of rotation (TAR), we determine whether rotation can both shift periods to an unusual range and reproduce the properties of the variability amplitudes. Furthermore, the centrifugal distortion induced by rotation can also affect luminosity and effective temperature of stars \citep{vonzeipel}. Hence the observed flux will depend on the stellar inclination \citep[e.g.][]{maederpeyt}. We determine for different rotational velocities and inclinations whether gravity darkening can displace SPBs towards fainter and cooler regions in the HR diagram.

The paper is structured as follows; in the first and second sections, we present respectively the stability analysis and visibility computations of two SPB models. In the third section, we determine how the visual properties of these models are affected by gravity darkening. The paper ends by a conclusion.

\section{Instability domain in the traditional approximation framework}
\label{Sec1}

The TAR is based on the assumption that the rotational frequency, $\Omega$, is moderate in comparison to the critical rotation rate of the star. Here, we assume a solid-body rotation and adopt $\Omega_{\textrm{crit}}=(GM/R^3_{\textrm{e}})^{1/2}$ as the critical rotation rate, where $G$ is the gravitational constant, $M$ the mass of the star, and $R_{\textrm{e}}$ the radius at the equator. 

The TAR is valid for high-order g mode and thus appropriate to study SPBs. \citet{townsend} showed that periods of classic SPB g modes can be significantly shifted with rotation whilst the limit of their instability strip is barely shifted towards cooler temperatures. In addition, \citet{savonije} and \citet{townsendretro} studied modes that have no counterpart in the case with no rotation. These modes are known as Yanai and Rossby waves and present a hybrid character, since their restoring force is a mix between the Coriolis force and buoyancy. The Yanai modes can present periods well below 1~d and could explain low period modes in late B-type stars as it was advanced by \citet{savonije}. However, these modes might be difficult to observe as we will show in Sec.~\ref{Sec2}. 

The limit of validity for the TAR is not clearly determined. \citet{ballottar} showed in a comparison with non-spherical models that the TAR underestimates the periods of modes up to 4\% for $\eta \simeq 5$, where $\eta=2\Omega/\omega_{\textrm{co}}$ is the spin parameter and $\omega_{\textrm{co}}$ the angular pulsation frequency. Since we look at effects that could shift periods from $\sim$1 to $\sim$0.3~d, that is by 70\%, the TAR computations should nevertheless remain of sufficient precision. 

\begin{table}
\begin{center}
\caption{Inertial periods (in days) of modes found to be excited in the 2.9 and 3.2 M$_{\odot}$ aged of $\sim$~20~Myr, and for various rotation frequencies. Figures between brackets give the number of excited modes.}
\label{Tab1}
\begin{tabular}{c|cccccc|}\hline 
\textbf{Mass} & \textbf{$\Omega/\Omega_{\textrm{crit}}$} & \textbf{$\ell=1$} & \textbf{$\ell=2$} & \textbf{$\ell=1$} & \textbf{Yanai} & \textbf{Yanai}\\ 
(M$_{\odot}$) &  & \textbf{$m=-1$} & \textbf{$m=-2$} & \textbf{$m=0$} & \textbf{$m=1$} & \textbf{$m=2$}\\ 
\hline
2.9 & 0.20 & 0.43 - 0.53 (7) & -- & -- & 5.09 - 8.63 (6) & 0.63 - 0.64 (8) \\
 & 0.40 & 0.29 - 0.33 (7) & -- & -- & -- & 0.298 - 0.304 (5)\\
 & 0.60 & 0.22 - 0.24 (7) & -- & -- & -- & --\\
3.2 & 0.20 & 0.43 - 0.58 (10) & 0.23 - 0.28 (8) & 0.57 - 0.84 (10) & 4.05 - 9.57 (11)  & 0.64 - 0.66 (10) \\
 & 0.40 & 0.30 - 0.35 (9) & 0.15 - 0.18 (9) & 0.46 - 0.58 (7) & 1.16 - 1.84 (9) & 0.30 - 0.32 (11) \\ 
 & 0.60 & 0.22 - 0.25 (9) & 0.12 - 0.13 (9) & 0.40 -0.45 (4) & 0.64 - 0.81 (7) & 0.19 - 0.20 (9) \\

\hline
\end{tabular}
\end{center}
\vspace{1mm}
\end{table}

With help of the CLES code \citep{cles} and adopting the solar chemical mixture, we compute two SPB models of 2.9 and 3.2~M$_{\odot}$, representative of the red border of the SPB instability strip. We perform a stability analysis with the MAD code including the TAR \citep{bouabid} for the models of about 20~Myr, in agreement with the age estimated for NGC~3766 \citep{aidelman}. Periods in the inertial frame ($P_{in}$) of the modes found to be excited are given in Table~\ref{Tab1}, where modes with azimuthal order $m<0$ are prograde. 

Among the classic SPB g modes, i.e. those presenting a counterpart in the case with no rotation, the prograde sectoral modes (PS; $m=-\ell$), are preferentially excited at the red border of the instability strip  \citep[see details in][]{townsend}. Moreover, their $P_{in}$ are shifted downwards 0.23 - 0.58~d to 0.12 - 0.25~d for 0.20 and 0.60~$\Omega_{\textrm{crit}}$, respectively. Axisymmetric modes ($m=0$) are only excited in the 3.2~M$_{\odot}$ case with periods around 0.5~d but tend to stabilise as $\Omega$ increases. The other classic g modes are not presented in Table~\ref{Tab1} because they are stable in almost every cases, excepting a low number of them in the 3.2~M$_{\odot}$ model rotating at 0.20~$\Omega_{\textrm{crit}}$ \citep[see more details in][]{salmon}.

Yanai modes with $m=1$ present periods in the usual SPB domain, though they decrease below 1~d as rotation increases (0.60~$\Omega_{\textrm{crit}}$). Clearly, $m=2$ Yanai modes present periods quite smaller than 1~d, as in the case of PS g modes.

\section{Visibilities of the unstable modes}
\label{Sec2}

Computing the amplitudes of non-radial pulsations is a difficult non-linear problem, which cannot be carried out presently. We instead compute mode visibilities that we normalise by the highest visibility reached by one of the modes. Initially suggested by \citet{savonije2}, this approach makes it possible for a relative comparison between the modes. 

\begin{figure}[b]
\begin{center}
\includegraphics[width=0.5\textwidth]{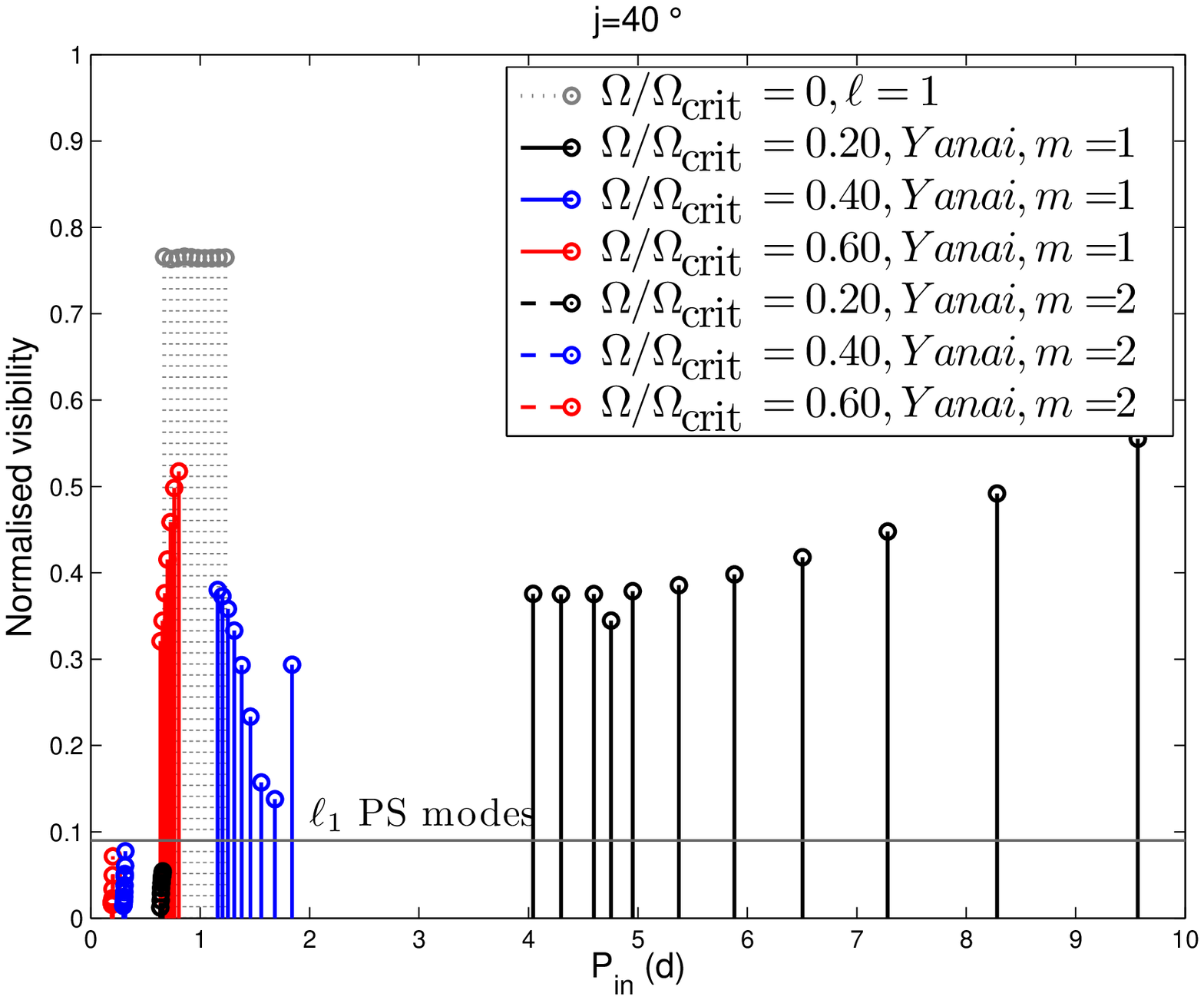}\includegraphics[width=0.5\textwidth]{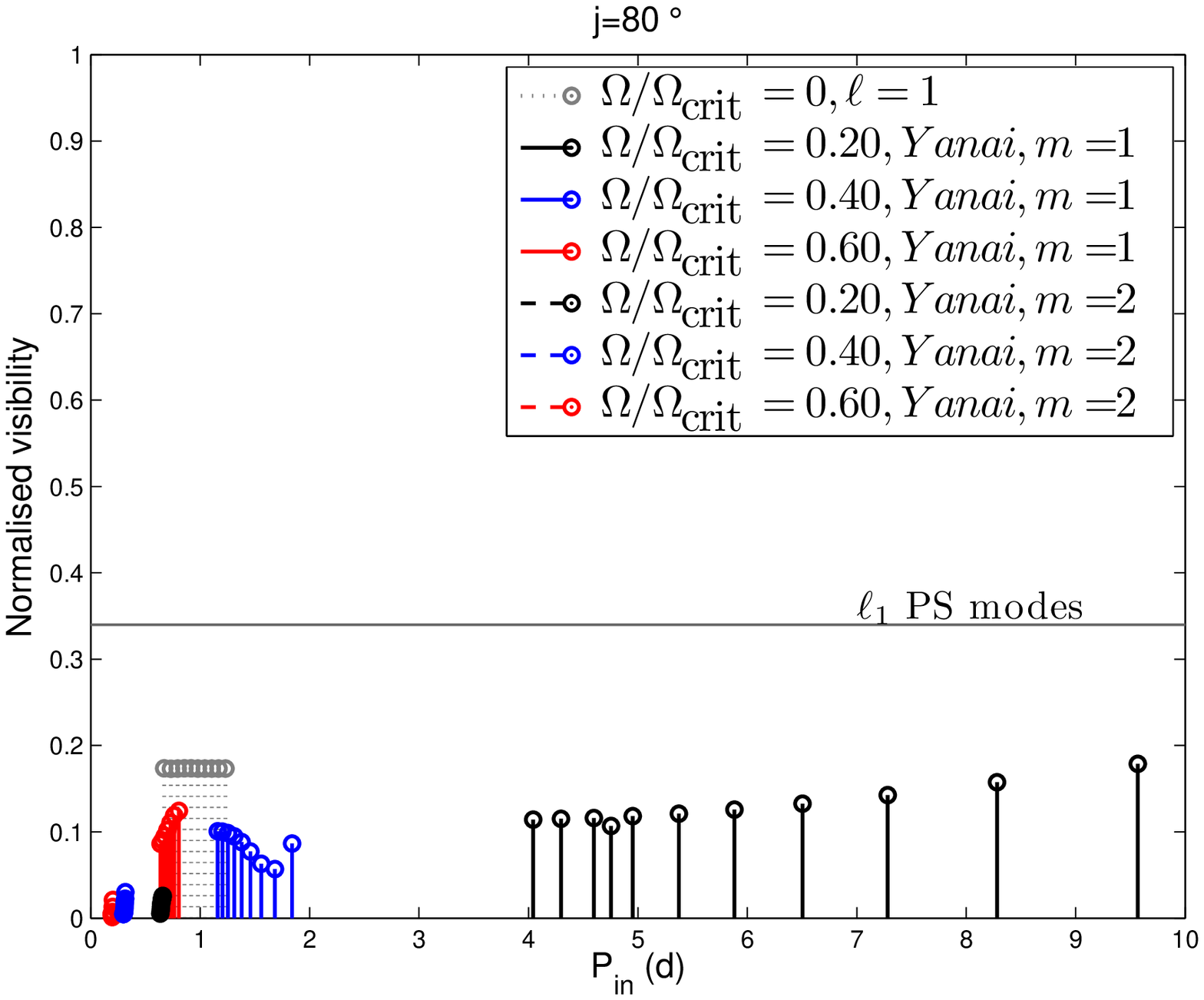} 
\caption{Normalised visibilities (see text) as a function of the inertial periods of all the Yanai modes found to be unstable in the 3.2~M$_{\odot}$ model aged $\sim$~20~Myr. Left (resp. right) panels correspond to an observer viewing angle of 40° (resp. 80°). The horizontal grey lines represent the normalised visibilities of the PS $\ell_1$ modes, which present the same value at 0.20, 0.40 and 0.60~$\Omega_{\textrm{crit}}$.}
\label{fig1}
\end{center}
\end{figure}

We determine the visibilities (in the CoRoT visible passband) following \citet{townsendvis} and adopting a normalisation as presented above \citep[see details in][]{salmon}. The visibilities of classic SPB g modes found to be excited in Table~\ref{Tab1} were presented in \citet{salmon}. We found that the PS $\ell=1$ ($\ell_1$ hereafter) modes are the most visible when the star is seen towards the equator, with visibilities typically $\sim$30-40~\% that of SPB modes in a case with no rotation. Therefore, the $\ell_1$ PS modes reproduce qualitatively the observed ratio between the amplitudes of the new variables and SPB stars of Mo13. 

The visibilities of the Yanai $m=1$ and $m=2$ modes are shown in Fig.~\ref{fig1}, where the horizontal lines depict those of the PS $\ell_1$ modes. As they present no equatorial confinement, Yanai modes are the most visible when the star is seen towards mid-latitudes ($j$=40$^{\circ}$), $m=1$ modes reaching up to 50~\% of the visibility of SPB modes with no rotation. However, their visibilities drop to $\sim$~20~\% and $\sim$~10~\% when the star is seen close to the pole ($j$=10$^{\circ}$, not presented) and close to the equator ($j$=80$^{\circ}$), respectively. In this latter case, their visibilities are $\sim$~3 times smaller than those of PS $\ell_1$ modes. The visibility of $m=2$ Yanai modes never exceeds 10~\%. As a consequence, Yanai modes with periods $<$~0.65~d should be difficult to detect in comparison with the PS modes. Hence, these latter appear as good candidates to explain both the periods and amplitudes of the new variable stars.

\section{Gravity darkening due to rotation}

We now determine whether the effect of rotation on the visual properties of SPB stars can make them appear outside their instability strip in the HR diagram. We assume the star is in solid-body rotation and adopt the Roche model to describe the centrifugal distortion. We then follow \citet{Georgy} to compute the effects of limb- and gravity darkening on the apparent effective temperature ($T_e$) and luminosity (L), depending on the inclination and rotation rate.  

\begin{figure}[b]
\begin{center}
 \includegraphics[width=0.55\textwidth]{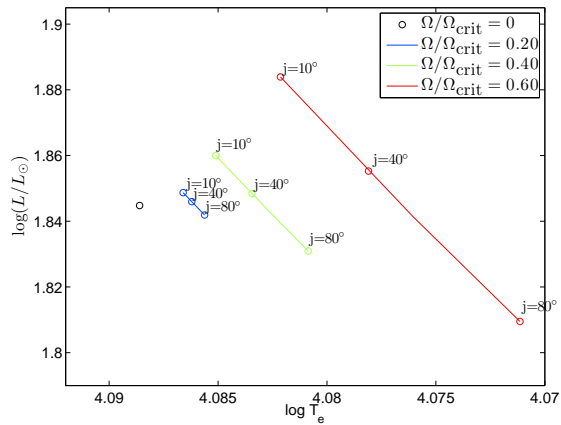}
\caption{Luminosity and effective temperature of the 2.9~M$_{\odot}$ model aged of $\sim$~20~Myr in the cases of no rotation and rotation at 0.20, 0.40 and 0.60~$\Omega_{\textrm{crit}}$, for inclinations $j$ from 10$^{\circ}$ to 80$^{\circ}$.}
\label{fig2}
\end{center}
\end{figure}

We illustrate these effects in Fig.~\ref{fig2} for a Geneva model \citep{genevacode} of 2.9~M$_{\odot}$ reproducing the properties of the CLES model with the same mass considered in the previous sections. The larger the rotation rate is, the cooler the stars appears. Meanwhile, the more the star is seen close to the equator ($j$=80$^{\circ}$), the more it appears cooler and fainter. It results that SPB stars at the red border of the instability strip can be displaced outside the strip and appears as if they were between the SPB and $\delta$ Scuti instability domains \citep[see][]{salmon}. In particular, the PS modes are the most visible when the star is highly inclined, i.e. when the star appears shifted at most towards low $T_e$ and $L$. 

\section{Conclusion}

The presence of fast rotators in NGC~3766 has led us to include rotation in our study, resulting in a scenario able to account for the variable stars detected in the cluster. Using the TAR, we have shown that PS modes of fast rotating SPB stars are good candidates to explain this new kind of variability. At the red border of the classic SPB instability strip, the periods of excited modes are shifted from 0.7--1.2~d (no rotation) to 0.15--0.6~d or 0.12--0.45~d depending on the rotation rate. In particular, the PS modes are there preferentially excited. They behave as Kelvin waves, which are confined to the equator, making them visible only in highly inclined stars (low latitudes facing the observer). 

Whereas, this combination of a fast rotation rate and high inclination coincide with large shifts towards cooler $T_e$ and fainter $L$ due to gravity darkening. It leads to a displacement of these pulsators outside the classic SPB instability strip. For such inclinations, the visibilities of $\ell_1$ PS modes are qualitatively 2 to 3 times lower than in a non-rotating star. These values correspond qualitatively to the ratio between the variability amplitudes of the SPB and new variable stars observed by Mo13. The Yanai modes with the lowest periods ($\lesssim$0.65~d, $m$=2) present low visibilities, clearly smaller than those of PS modes. However, some of the new variable stars close to the SPBs detected by Mo13 present modes of both SPB- and new variability type; they might correspond to stars less inclined and presenting $m$=1 Yanai or axisymmetric classic g modes.

This scenario might also explain the similar new variable stars detected by CoRoT \citep{degrootecorot}, and in the NGC~884 and NGC~457 clusters \citep{saesenpersei,polonais}. In the future, spectroscopic campaigns devoted to these stars could in particular confirm the role of rotation.

\bibliographystyle{iau307}
\bibliography{biblio}

%
%
%
%
%

\begin{discussion}

\discuss{Meynet}{In case faster rotating stars are more frequent at low metallicity, can we expect more stars showing this type of variability?}

\discuss{Salmon}{I expect, for a lower metallicity, a lower number of prograde sectoral modes found to be excited, since their driving mechanism is dependent on the metal content of the star. Similarly, excited axisymmetric modes should only appear to be excited in slightly more massive SPB stars than at larger metallicity.}

\discuss{Anderson}{Since you argue that the new variables in NGC 3766 are seen equator-on, have you deduced whether this is supported by random inclinations in that sample of stars?}

\discuss{Salmon}{We plan in the future to carry out kinds of Monte-Carlo simulations of stellar populations in which the rotation rates and inclinations would vary randomly. We would then check whether the fraction of stars in these simulations that fits the conditions to present prograde sectoral modes agrees with the observations. Yet, the fact that some new variable stars close to the SPBs detected in NGC 3766 present modes with periods both close to 1 day and 0.3 day could correspond to less-slanted stars, showing axisymmetric or Yanai modes and being hence less shifted due to the gravity darkening.}

\discuss{Saio}{If the traditional approximation is not used, some modes are damped due to mode interactions.}

\discuss{Salmon}{I agree with that comment. Lee (2008) computed pulsations in a 2D model, taking into account centrifugal distortion and showed that the driving of retrograde modes was less efficient, in part due to coupling between modes. On the contrary, the driving of prograde modes seemed to be amplified.}

\end{discussion}

\end{document}